\documentclass[aps,prl,twocolumn,showpacs,floatfix,superscriptaddress]{revtex4}

\usepackage{graphicx}  % needed for figures
\usepackage{dcolumn}   % needed for some tables
\usepackage{bm}        % for math
\usepackage{amssymb}   % for math
\usepackage[latin1]{inputenc}
\usepackage{amsfonts}
\usepackage{amsmath}
\usepackage{amsthm}
\usepackage[american,english]{babel}
\usepackage{fontenc}
\usepackage{textcomp}
\begin{document}

%\title{When {\em Big Data} Fails! Information coarse-graining and the
%relative success of adaptive agents competing for a limited resource}
%\title{When {\em Big Data} Fails! Relative success of agents using
%coarse-grained information in complex adaptive systems}
\title{When {\em Big Data} Fails! Relative success of adaptive agents
using coarse-grained information to compete for limited resources}

\author{V. Sasidevan}
\email{sasidevan@gmail.com}
\affiliation{The Institute of Mathematical Sciences, CIT Campus, Taramani, Chennai
600113, India.}

\author{Appilineni Kushal}
\email{akushalstar@gmail.com}
\affiliation{Indian Institute of Science, C V Raman
Road, Bangalore 560012, India.}

\author{Sitabhra Sinha}
\email{sitabhra@imsc.res.in}
\affiliation{The Institute of Mathematical Sciences,
CIT Campus, Taramani, Chennai 600113, India.}

\date{\today}
%
% Use the package "url.sty" to avoid
% problems with special characters
% used in your e-mail or web address
%

\begin{abstract} 
The recent trend for acquiring big data assumes that possessing
quantitatively more and qualitatively finer data necessarily provides
an advantage that may be critical in competitive situations. Using a
model complex adaptive system where agents compete for a limited
resource using information coarse-grained to different levels, we show
that agents having access to more and better data can perform worse
than others in certain situations. The relation between information
asymmetry and individual payoffs is seen to be complex, depending on
the composition of the population of competing agents.
\end{abstract}
%02.50.Le Decision theory and game theory
%89.75.-k Complex systems
%87.23.Ge Dynamics of social systems
\maketitle 
%\section{Introduction}
%\label{sec:1}
Agents in a population often coordinate their actions with that of their
neighbors resulting in striking forms, such as in swarming and
flocking~\cite{Castellano2009,Vanni2011}.
%Patterns arising from individual entities in a population, coordinating 
%their actions with that of their neighbors, can often result in
%striking forms including swarming and
%flocking~\cite{Castellano2009,Vanni2011}. 
Typically, in
such cases, individuals use information obtained from their local
environment to adjust their actions in order to achieve some desired
objective~\cite{Nowak1992,Couzin2007,Pearce2014,Pinheiro2016}.
%e.g., as in spatial games~\cite{Nowak1992}. 
Emergent coordination is therefore crucially dependent on the
information acquired by an agent and its ability to process it
appropriately, which determines its future course of action.
Often the objectives of different agents in a system may not be compatible with
each other, for instance when they are competing for a limited
resource. Examples of such situations 
%in which agents have mutually opposed objectives 
are abundant in nature, where individuals vie for food,
shelter and mating opportunities. Even in our more complex social
environment, we regularly come across instances of such competition,
e.g., people trying to choose the least congested route through an urban
road network or anticipating whether the relative demand for a
financial asset will increase,
so as to profit by buying or selling it 
%betting appropriately
at the present~\cite{Challet2001}.
%In socio-economic settings we regularly come across individuals
%trying to find the least congested road or path in a network, trying
%to beat the market., anticipate a less congested compartment in a
%train, less congested time in the institute canteen etc. 
In these settings, individuals may use
%In such complex situations, agents may use 
strategies which project
information about past experiences to make decisions about the
future course of action~\cite{Simon1955,Arthur1994,Arthur1999}.
Conventional wisdom suggests that the relative success of an agent in
meeting its objective (i.e., gaining access to the scarce resource)
would increase with the quality and quantity of
available data that would form the basis for its decisions. Indeed the
recent excitement about ``big data'' is partially based on the premise that
access to more and better information provides a
competitive advantage~\cite{Mattmann2013}.
%can be used to infer the consequences of the
%previous actions chosen by it and its neighbors. 
%Making use of information (data) gathered from past experiences
%and projecting them into the future to make decisions,  help the agents to
%simplify the decision making process in a complex environment. It is
%obvious that the efficacy of such decision making could depend upon
%the quality and/or quantity of data from the past available to the
%agents and how much is their processing capacity. In fact the new buzz
%word of big-data suggests that we should obtain more and more fine
%quality data to understand the complex dapative systems around us in a
%better way. 

In this paper we show that agents using quantitatively more data that
is also finely resolved (and hence also qualitatively superior) may
not actually do better - and can in fact lead to significantly worse
payoffs - in situations where they are competing with
agents that have access to less, as well as more coarse-grained,
information.
This surprising result arises from
emergent coordination in the collective activity of agents who use
information of a particular quality (i.e., level of coarse-graining)
leading to macroscopic patterns of behavior that may be discernible
from the data only at a different level of coarse-graining. Thus, if
there are other agents in the population who have access to
information about the system at this latter coarse-grained level, they
can potentially exploit this predictability to their advantage. We
show this using a 
%model complex adaptive system comprising agents,
%distinguished in terms of the quality and quantity of information that
%they use to choose between a number of possible options in order to
%gain a higher payoff, viz., preferential access to a limited resource.
model of preferential access to a limited resource.
This comprises a complex adaptive system of agents, each attempting to
maximize their payoffs. The agents are distinguished in terms of the
quality and quantity of information that they use to choose between
several possible options.
%SPLIT THE SENTENCE BELOW
%In this setting, one can vary
%By varying the composition of a population comprising 
The setting allows us to vary
the composition of the population in terms of the different types of
agents, each of whom exclusively uses one of two types of historical
data about the system that represent the two extremes of
coarse-graining. We show that
the relation between information asymmetry and the performance of
agents is a complex one, depending on the relative fraction of the
population that each type of agents constitute. Thus, the utility of
``big-data'' is contingent upon the precise nature of the ecosystem
comprising all its competitors that an agent is located in. 
%SPLIT THE SENTENCE BELOW
The premise that more and better information will automatically result in
better performance, e.g., by improving predictive power, therefore
needs to be treated with caution. This is especially true for competitive
situations where adaptation through learning is involved that are
ubiquitous in systems around us, such as financial
markets~\cite{Potters2007,Sinha2010}.

%\section{The Model}
%\label{sec2}
%MG. Detailed Definition. Binary information Vs exact info. recap of
%results. our study varies the population composition.
To investigate how information asymmetry between agents affect their
performance, specifically
when different agents use information at diverse levels of
coarse-graining,
we focus on a complex adaptive
system where agents compete for a limited resource. 
Here the heterogeneous agents
%(in terms of the information they have access to) 
use the different types of information that they have access to
for the same
%use the {\em different} information they have access to for the same
purpose, viz., to have preferential access to the resource.
In particular, we use the paradigm of the Minority Game
(MG)~\cite{Challet1997,Moro2004,Challet2005} which has
all the ingredients to address the above
question in a quantitative manner. In addition, it has the advantage
that the classical version, in which agents use information only at 
a single level of coarse-graining, is well-understood and 
can be used as a benchmark for the more complex situation that is
investigated here.
We consider a population of an odd number $N$ of agents who
independently and simultaneously choose between two options ($A$ and
$B$, say) in each round. 
The option that is chosen by fewer agents is considered the better
choice (outcome) in each round and leads to a higher payoff (say, 1),
while those who had chosen the alternative
%other option
receive a lower payoff (say, 0). 

\begin{figure}[tbp]
\includegraphics[width=.9\linewidth]{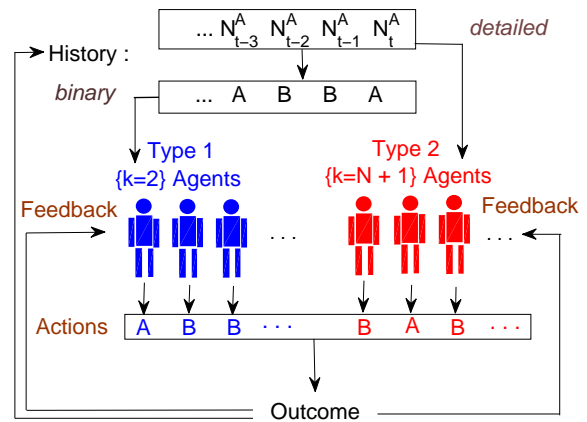}
\caption{A schematic representation of a 
%minority game involving
%resource allocation game 
complex adaptive system comprising $N$ agents that are competing for a
limited resource. Every agent has to choose between two possible
actions ($A$ or $B$) at each round, with the option chosen by the
lesser number of agents being the better choice (outcome) in that round.
The agents decide on their choice using strategies based on
information about the previous $m$ rounds' collective choice dynamics,
which could result in the system being in one of $k$ possible states ($2 \leq k
\leq N+1$, depending on the level of coarse-graining) at each round.
Here agents have been distinguished into two classes (Types 1 and 2)
according to the two extreme levels of coarse-grained information,
i.e., $k=2$ and $k=N+1$, respectively, that they have
access to.
%Type 1 agents use binary information 
%of two possible types, distinguished by the granularity of
%information, viz., binary and detailed, about the history of outcomes
%that they have access to and which they
%use to make their choice in the next round. 
%Binary information ($k=2$) agents use the identity of which of the two
%options were chosen by the minority in the previous rounds, while
%detailed information ($k=N+1$) agents use the exact number
%of agents $N^A$ who chose for a particular option A.
After each round $t$, the detailed information about the total number of
agents choosing a specific action $A$ (say),
$N^A_{t}$,
that is accessed only by Type 2 agents, as well as, binary information, viz.,
the choice of the minority ($A$ or $B$) which is accessed only by Type 1 agents, 
are added to the history of outcomes. The information about the
outcome is also used as feedback for adaptive selection of strategies by the
agents.
}
\label{fig:1}       % Give a unique label
\end{figure}

We assume that the population consists of different {\em types} of
agents distinguished in terms of the data that they have access to,
which is coarse-grained to different levels (Fig.~\ref{fig:1}). 
The level of coarse-graining $k$
($2 \leq k \leq N+1$) is defined in terms of the extent to which an
agent can resolve the number of agents ($N_A$, say) opting for a
particular choice ($A$).
%distinguished in terms of the level of coarse-graining in the
%data about past events that they have exclusive access to.
For example, an agent with $k=2$, the coarsest level of resolution,
can only distinguish between $N_A >
N/2$ and $N_A < N/2$ (i.e., whether $A$ was chosen by the majority or
not) in a particular round~\cite{Challet1997}. Conversely, the finest level of
resolution corresponds to $k=N+1$, for which an agent can determine the
exact number of agents $N_A$ opting for $A$ in a round~\cite{sasidevan}.
Any value of $k$ between these two extremes represents intermediate
levels of coarse-graining where the agent can only distinguish
whether, in each round, 
$N_A$ belongs to any one of the intervals $[i N/k, (i+1)N/k]$ where
$i=0, \ldots, k-1$.
%For e.g a $k=3$ agent can only distinguish whether $N_A$ is
%between $i N/3$ and $(i+1)N/3$ where $i = 0,1,2$.
The agents of each type use the appropriately coarse-grained data to
determine their choice of action in the next round.
%We emphasize that all these agents use their respective coarse grained
%data for the same purpose.i.e, to predict the next minority side ( a
%binary prediction). In this respect agents use strategy tables as in
%the classical MG setting which project past coarse grained data to
%future action. 
For clarity we focus on the interaction between only two types of agents
corresponding to the extremes of coarse-graining, viz., $k=2$
(which we designate as Type 1 agents) and $k=N+1$ (Type 2 agents).
%The number of preceding rounds whose information that each type of
%agent retains is referred to as their memory length, 
The memory length of each type of agent indicates the number of past
rounds whose information they retain,
and is denoted by $m_1$ ($m_2$) for Type 1 (Type 2) agent.
%%%%%MOVE FOLLOWING TO SI%%%
Each agent uses {\em strategies} that map the information about
past events ($m_1$ bits for Type 1 agent, $m_2 \log_2 (N+1)$ bits for
Type 2 agent) to the choice of action in the next round (i.e., $A$ or
$B$).  Each agent initially chooses at random a small sample of
strategies (e.g., of size 2 as here) from the set of all possible
strategies, which is of size $2^{2^{m_1}}$ for a Type 1 agent and
$2^{(N+1)^{m_2}}$ for a Type 2 agent.  At each round, an agent scores
the strategies according to the potential payoffs that would have been 
obtained by using
them in the previous rounds (feedback), and uses the one having the
highest score.
%%%%%UPTO HERE %%%%%%%%%%

%NEW
%In order to examine the role of information coarse-graining in complex
%adaptive systems, 
%NEW

%It is known that in the conventional MG, agents having large enough
%memory size self-organize into a state where the fluctuation in the
%number choosing a particular option about the mean value ($N/2$ due
%to the symmetry between A and B) is minimized. This results in the
%population as a whole doing better (i.e., it is globally efficient)
%than the case in which agents choose randomly between A and B with
%equal probability.  The most globally efficient state is achieved for
%a critical value of memory size of agents, $m_c \sim \log_2
%N$~\cite{challet1998,challet2000,challet2005}.  When the memory size
%is $\ll m_c$, the agents exhibit herding behavior where most of them
%choose the same option in a given round, resulting in very large
%fluctuations about the mean. For such a situation, the individual
%payoffs are extremely low and the system is also globally inefficient
%- even compared to simple random choice behavior.

\begin{figure}[tbp]
\includegraphics[width=.99\linewidth]{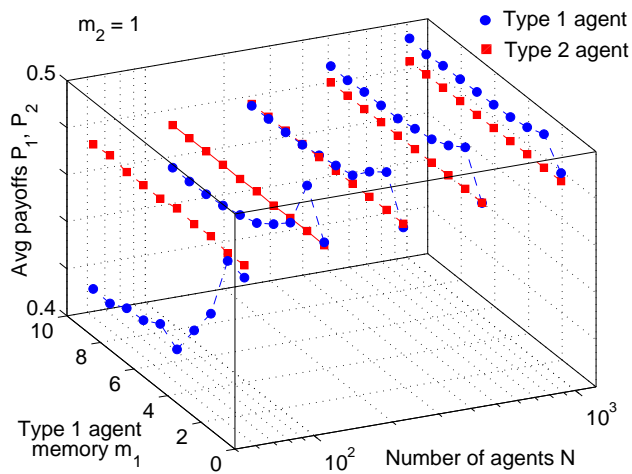}
\caption{Average payoffs $P_1$, $P_2$ of Type~1 and Type~2 agents
(respectively) shown as a function of
the memory length $m_1$ of a single Type 1 agent interacting with
$N-1$ Type 2 agents with memory length $m_2 =
1$ for different population sizes $N$.
Note that when the lone
Type~1 agent has a lower memory length $m_1$, it 
has a relative advantage over Type~2 agents, 
receiving the highest payoff when $m_1 = 2$.
%The variation of the payoffs with $m_1$ shows different profiles for
%different memory lengths (a) $m_2 = 1$ and (b) $m_2=2$ of the Type~2
%agents. In both cases, the Type~1 agent receives the highest payoff
%when its memory length $m_1 = 2$. Note that, the lone Type~1 agent
%has a relative advantage over Type~2 agents, when the former has a
%lower memory ($m_1$).
Payoffs are averaged over $10^4$ iterations in
the steady state and over 100 different realizations.
}
\label{fig:3}
\end{figure}

%\section{Results}
%\label{sec3}
In order to study how the relative performance of agents in choosing their
optimal future action is affected when different agents have access to
data with different levels of coarse-graining (and therefore,
qualitatively distinct information), we first focus on the simplest
case of a {\em single} Type 1 agent with memory length $m_1$ interacting
with a population of $N-1$ Type 2
agents with memory length $m_2$.
Note that both types of agents use the different information available
to them (representing the two extremes of coarse-graining) with the
identical aim of predicting the outcome in the next round.
One may naively expect that agent(s) having more information at their
disposal (e.g., as measured in units of bits) will have
an advantage over the other type of agents. Consequently, it would
have been expected that when the number of bits, $m_1$, in the information
accessible to the Type 1 agents is less than $m_2 {\rm log}_2 (N+1)$, the
corresponding quantity for the Type 2 agents, then the latter would
have obtained a relatively higher payoff. This would also be in
accordance with the intuitive notion that the highly resolved detailed
data of Type 2 agents is qualitatively better than the low-resolution
outcome data of Type 1 agents. However, the mean payoffs of the two types of
agents shown in Fig.~\ref{fig:3} for different memory lengths
$m_1$
and population sizes $N$ reveals that the actual behavior is
more complex.

The most surprising outcome for the case when the Type 2 agents have
memory length $m_2 = 1$ (Fig.~\ref{fig:3}) is that the Type 1
agent is able to acquire a relatively higher payoff at low values of
$m_1$ even though the information accessible to it is highly
coarse-grained and quantitatively much less (in terms of bits) compared to
the rest of the population. Moreover, the range of $m_1$ over which
the Type 1 agent does better than Type 2 agents is seen to increase
with $N$. Thus, the success of an agent in a complex adaptive system, where the
information accessible by the individual entities differ both in
terms of quality and quantity, is not entirely determined by the amount and
resolution of the data at
its disposal. Instead, as we show below, it depends more on whether discernible 
patterns in the behavior of the population are present at the level of
coarse-graining it has access to.
%Note that the Type1 agent with $m1 = 1$ don't have any advantage. This
%can be understood based on the symmetry between the two options A and
%B. It is expected that both A and B will be on the minority side equal
%number of times. So A and B should occur with equal probability
%following an A or B in the time series. This implies that a single
%memory $m1 = 1$ Type1 agent is as good as a randomly choosing player
%and will not have any advantages of adaptive players. 
%
%MOVE TO SUPPLEMENTARY INFO %%%%%%%%%%%%%%%%%%
When the memory length of the Type 2 agents is increased to $m_2=2$
(see Supplementary Information), the Type 1 agent is no longer
observed to have a higher payoff than the rest of the population,
regardless of its memory length $m_1$. Note that Type 2 agents attain
the highest degree of emergent coordination among themselves for $m_2
= 2$ independent of $N$~\cite{sasidevan}. Thus, it is not surprising
that the lone Type 1 agent will not be able to outperform the
optimally coordinated population of Type 2 agents.  However, as we
shall show below, introducing multiple Type 1 agents makes it
possible for these mutually competing individuals to develop emergent
coordination within themselves by which they can outperform Type
2 agents with $m_2 = 2$.  If $m_2>2$, the behavior of the Type 2
agents is indistinguishable from agents randomly choosing between $A$
and $B$~\cite{sasidevan}. 
As there is no predictability in the time-series available
to the Type 1 agent that it can exploit, it will on average receive
essentially the same payoff as the rest of the population.
%%%%%%%%% UPTO HERE %%%%%%%%

The opposite case of a single Type 2 agent interacting with a
population of $N-1$ Type 1 agents shows that the lone Type 2
agent will perform better than the rest of the population when $m_1$
is low, independent of $m_2$ (see Supplementary Information). 
This can be explained from the
observation that at low $m_1$, the behavior of a population of Type 1
agents has a periodic pattern in the number choosing a
specific option - which is precisely the information accessible by the
Type 2 agent. It is therefore optimally placed to exploit the
predictability to its advantage.
%Indeed this is what we observe
%for $m2=2$ from numerical studies as described later. 
%
%Note that in Fig.\ref{}, we considered the case where $m2 = 1$. Now
%what happens when $m2$ is increased..? We already know that at $m2=2$,
%the Type 2 agents achieve maximal emergent coordination. So we expect
%that the lone Type 1 agent will  not have any advantage in this case,
%which is indeed what we see from simulation studies. For the cases
%where $m2 > 2$, the Type 2 agents essentially behaves like randomly
%choosing agents. We expect some advantage for Type 1 agents in such a
%case but only when there is a sufficient number of them to have
%coordination effects to come into play.

%\subsection{Information and predictability}
To explain the relative performance of different types of agents
having access to information at the two extreme 
levels of coarse-graining, we focus on
the information content in the history of outcomes 
that can be exploited by the agents to their advantage. 
As we shall see, the collective action of any one type of agents may result in
predictable patterns at the other level of coarse-graining and hence
observable only to these other agents.
This ``useful'' information content above the noise level 
can be quantified by measuring the predictability of a particular
choice (say $A$) being the outcome in a particular round, given the 
history of past outcomes. This history can be either the {\em
binary} sequence of outcomes $A$,$B$
or the {\em detailed}
time-series of the number of agents $\{N^{A}_{t}\}$ choosing a
particular option $A$, the former (latter) being accessible only to a
Type 1 (Type 2) agent. We can therefore define two distinct
information measures, viz., $H_1 = \sum_{u_{L_{bin}}} P(u_{L_{bin}})
[P(A|u_{L_{bin}}) - (1/2)]^2$, and $H_2 = \sum_{u_{L_{det}}}
P(u_{L_{det}}) [P(A|u_{L_{det}}) - (1/2)]^2$.
%\begin{align}
%H_1  & = 
%\sum_{u_{L_{bin}}} P(u_{L_{bin}}) [P(A|u_{L_{bin}}) - (1/2)]^2,\\
%{\rm and}~~H_2  & = 
%\sum_{u_{L_{det}}} P(u_{L_{det}}) [P(A|u_{L_{det}}) - (1/2)]^2.
%\end{align}
Here, $u_{L_{bin}}$ is the binary sequence of
outcomes for the previous $L_{bin}$ rounds 
while $u_{L_{det}}$ is the sequence of integers, each lying between $0$ and
$N$, representing the number of agents choosing $A$
in the previous $L_{det}$ rounds.
%(The $|~.~|$ refers to the cardinality, i.e., the total number of
%possibilities, of these sequences.
The probability with which a particular sequence of $L$
successive outcomes is observed is denoted as $P(u_{L})$, while 
$P(A|u_{L})$ represents the conditional
probability that the outcome $A$ follows the sequence $u_{L}$.
%Note that, while for a population of Type 1 agents, each
%possible outcome sequence 
%of length $L$ occurs with equal probability 
%when $L \leq \log_2 N$~\cite{cavagna_1999,challet_2000},
%for a population of Type 2 agents different histories of length $L$
%occur according to a Gaussian distribution~\cite{sasidevan}.

\begin{figure}[tbp]
\includegraphics[width=.99\linewidth]{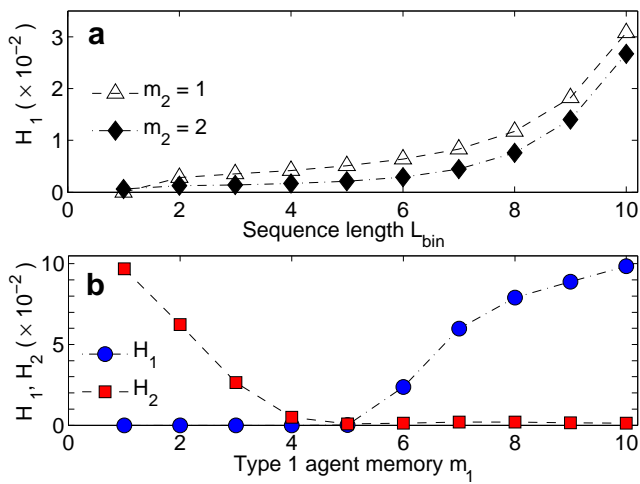}
\caption{(a) Information content $H_{1}$ of the binary sequence
containing the history of outcomes for a game involving only Type~2
agents shown as a function of the sequence length
$L_{bin}$. (b) Information content of the binary sequence, $H_1$,
and that of the time series containing detailed information (exact
number opting for a particular choice) of sequence length $L_{det}
=1$, $H_2$, for a game involving only Type~1
agents, shown as a function of
the memory length $m_1$ of the agents. 
The number of agents considered in both (a) and (b) is $N=255$.
Results shown are averaged over 100 realizations.}
\label{fig:4}
\end{figure}

We first consider the case of a population comprising only Type 2
agents having memory length $m_2$. The collective behavior of such
agents generates a history of {\em binary} outcomes whose information
content $H_1$ is shown in Fig.~\ref{fig:4}~(a) for $m_2 = 1$ and $2$. 
Note that this
information cannot be used by the Type 2 agents themselves, whose
strategies are based on $u_{L_{det}}$ but is accessible in principle to
a hypothetical Type 1 agent whose strategies use $u_{L_{bin}}$.
%%The case of a pure population of Type 2 agents is more remarkable
We observe that $H_1$ increases with the length of the binary
sequence, $L_{bin}$, over the range of sequence lengths
considered here, with $H_1=0$ when the
history is restricted to the immediately preceding round, i.e., $L_{bin}
=1$. 
%The symmetry between the two options A and B imply that there is no
%predictability when $L_{bin}= 1$. 
Thus, if a Type 1 agent with memory length $m_1$ is introduced into this
population, it can make use of the predictability present in the 
binary sequence accessible to it when $m_1>1$.
As $m_1$ increases, the
number of possible strategies that can be used by the Type 1 agent increases
exponentially ($= 2^{2^{m_1}}$). It therefore becomes progressively less likely
that the agent will randomly pick the strategy that can optimally exploit
the predictability present in $u_{L_{bin}}$.
This implies that the highest payoff of Type 1 agent is achieved for
the lowest value of $m_1$ having non-zero information content, i.e., 
$m_1 = 2$, as is indeed confirmed by Fig~\ref{fig:3}.
%confirms this by showing that the payoff
%obtained by a single Type 1 agent interacting with a population of Type 2
%agents is highest for $m_1 = 2$.
%resulting from the collective behavior of Type 2 agents.
%To make use of the predictability, the binary string representing a
%strategy must be sufficiently skewed in the distribution of its 0s and 1s.
%This explains why the CZMG agent has a relatively better performance
%in terms of payoffs despite using quantitatively lower amount of data
%than the DIMG agents.

We next consider the other extreme case represented by a population
comprising only Type 1 agents having memory length $m_1$.
This situation has been studied earlier in the context of
understanding how the collective behavior of such agents undergoes a
phase transition as $m_1$ is varied~\cite{challet1998,challet2000}, where the focus is on the
information content $H_1$ of the binary
sequence of outcomes available to these
agents~\cite{challet1999,hui1999,challet2000_1,challet2000}.
Here our focus is instead on the information content $H_2$ of the
detailed
time-series $u_{L_{det}}$ recording
the number of agents choosing a particular option (shown in
Fig.~\ref{fig:4}~(b) for $L_{det} = 1$).
It is important to note that the
latter information cannot be used by the Type 1 agents whose
interactions produce it, as their
strategies are based on $u_{L_{bin}}$. However, it is accessible to a
hypothetical
Type 2 agent with an appropriate memory length, viz., $m_2 =
L_{det}$. 
We observe that $H_2$ is non-zero even when $H_1 = 0$ 
%at low values of the memory length $m_1 \lesssim {\rm log}_2 N$.
at low values of the memory length $m_1$.
This indicates that the detailed history $u_{L_{det}}$ contains
potential predictability that can be exploited by a Type 2 agent.
Thus, both cases of
homogeneous agent type populations
considered above demonstrates that the
collective behavior of agents having access to information
coarse-grained at a specific level can give rise to some amount of
predictability that can be perceived only in the information available at 
a different level of coarse-graining. 
%We note in passing that the variation of the information content $H_1$
%available to the Type 1 agents having memory length $m_1$ reproduces
%the well-known phase-transition behavior (around $m_1 \sim log_2 N$)
%of the conventional Minority Game~\cite{}. 
%
%This provides an explanation
%of the relatively better performance of a single Type 2 agent interacting
%with $N-1$ Type 1 agents for lower values of $m_1$ as shown in
%Fig.~\ref{fig:2}.

%\subsection{Varying the composition of agent types in a population}
The above arguments explain the performance of a single agent of
one type interacting with a population consisting exclusively 
of agents of the other type. However, in reality, the number of each
type of agents
having access to data at different levels of coarse-graining can be
arbitrary. We shall now, therefore, consider the situation where the
relative fraction of the two types of agents present in the population
is varied 
%to span the interval 
between the two extreme cases
considered earlier.
%We now consider the situation when the
%composition of a population in terms of agents having access to binary
%and detailed information is varied between the two extreme cases
%discussed above. 
The important effect of introducing more than one agent of a specific
type into the population is that effective coordination between these
agents can emerge, resulting in enhanced payoffs for them. 
%It should be intuitively clear that introducing
%multiple CZMG agents in a population of DIMG agents may lead to the
%few CZMG agents using the information accessible to them in order to
%coordinate their actions and thereby increase their payoff.
%Conversely, introducing multiple DIMG agents in a population of CZMG
%agents could result in a higher payoff for the few DIMG agents.
\begin{figure}[tbp]
\includegraphics[width=.99\linewidth]{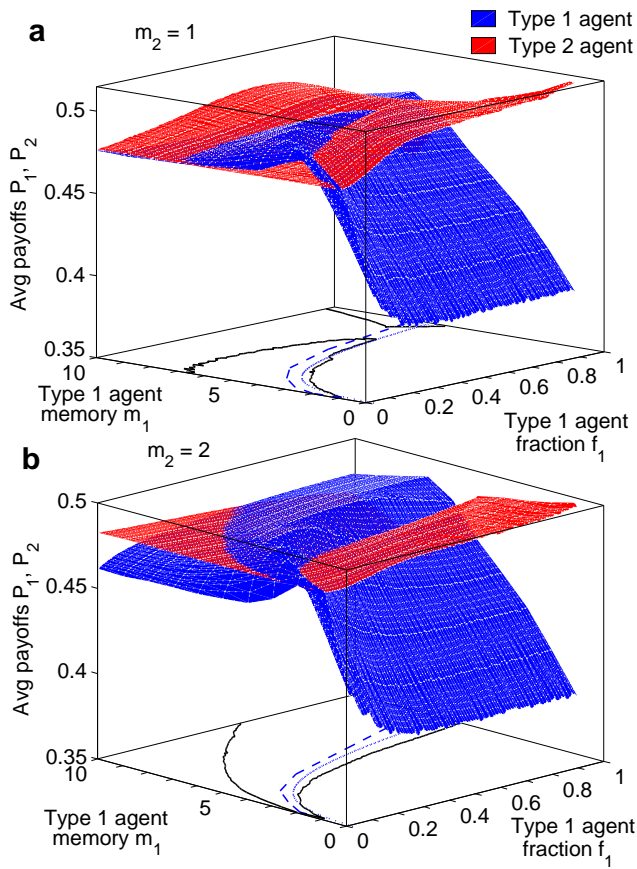}
\caption{
The average payoffs $P_1, P_2$ of Type 1 (shown in blue) and Type 2 agents (red)
comprising a population of size $N (=255)$
for different population fractions $f_1$ and memory length $m_1$ of
the Type 1 agents. The memory length of the
Type 2 agents are fixed at $m_2$ [$= 1$ for (a) and $=2$ for (b)].
The contours separate the regions in the ($m_1, f_1$) parameter 
space where Type~1 agents have a relative advantage over Type 2 agents
and vice versa. The broken curve represents the optimal population fraction
$f_1^*$ of Type 1 agents with a given memory length $m_1$ at which
they receive the highest payoff. The dotted curve is the value of
$m_1$ at which $N f_1$ Type 1 agents are expected to have maximum
payoff in absence of any Type 2 agents.
Payoffs are averaged over $10^4$ iterations in
the steady state and over 100 different realizations.
}
\label{fig:5}       % Give a unique label
\end{figure}

We gradually vary the fraction
$f_1$ of Type 1 agents in a population of total size $N$ comprising
agents of both Types 1 and 2. 
Fig.~\ref{fig:5} shows the average payoffs $P_1, P_2$ for agents of
each type respectively, for
different values of $f_1$ and memory length $m_1$ of Type 1 agents,
keeping the memory length of Type 2 agents fixed [viz., $m_2 = 1$ (a)
and $=2$ (b)]. 
As shown earlier for the case of a single Type 1 agent in a population of Type 2
agents, here also we see that Type 1 agents can
outperform Type 2 agents even when the quantity of information
(measured in bits) available to the former is much less than that for
the latter. This is particularly striking when the memory length $m_1$
of Type 1 agents is low, i.e., information content of each is $m_1$
bits $\ll m_2 {\rm log}_2 (N+1)$ bits, which is the information content of a
Type 2 agent. Fig.~\ref{fig:5} shows that, in this low $m_1$ regime, Type
1 agents when present in small numbers (i.e., low $f_1$) can receive
higher payoffs than the
Type 2 agents who form the bulk of the population.
We expect the situation that we considered earlier, viz.,
a single Type 1 agent playing against $N-1$ Type 2 agents, to occur
when $f_1 \rightarrow 0$. Thus, we expect the Type 1 agent to achieve
best performance (i.e., maximum payoff) for a memory size $m_1^* = 2$
which is indeed observed. On the other hand, as $f_1 \rightarrow 1$
the setting is that of a conventional MG between Type 1 agents
where the maximum payoff occurs for $m_1^* \simeq {\rm log}_2 (0.337
N)$~\cite{challet2000} which is also seen. Indeed, for any fraction $0 < f_1 <
1$ of Type 1 agents, their best performance is achieved for a memory
size $m_1^*$ that lies between $2$ and ${\rm log}_2 (0.337 N)$
[indicated by the broken curves in Fig.~\ref{fig:5}].
Thus, having multiple Type 1 agents in the population can help them
achieve a higher payoff than they are capable of by playing singly
against a population of Type 2 agents, suggesting an important role of
emergent coordination among a group of competing agents who are 
distinguished by the nature of information available to them.
%We observe that when $m_1$ is low, Type 1 agents have an advantage
%over the other type of agents when they are present in extremely small
%numbers (i.e., $f_1$ is low). 

A simple qualitative argument for this locus of maximum
payoffs for Type 1 agents in the ($m_1,f_1$) parameter space is as
follows. Ignoring for a moment the presence of Type 2 agents, we can
consider the population to exclusively comprise $f_1 N$ Type 1 agents
with memory size $m_1$.
This population will achieve their highest payoff at
$m_1 \approx {\rm log}_2 (0.337 N f_1)$ [for $f_1 > 3/N$, so that $m_1
>0$] shown by the dotted curves in
Fig.~\ref{fig:5}. The presence of Type 2 agents results in a shift
towards lower values of the optimal population fraction $f_1^*$ of
Type 1 agents
(as shown by the empirically
obtained curves of maximum payoff of Type 1 agents represented by the
broken curves). Note that this argument does not say anything about
the relative performance of the two types of agents. The relatively
higher payoff of Type 1 agent (with low $m_1$) compared to Type 2
agents, despite the latter
having quantitatively more information for decision-making, is thus an extremely
surprising outcome that emerges from the collective dynamics of
interactions between agents with access to information coarse-grained
at different levels.
%Beyond $m1 = 0.337 \log_2 N$, the payoff for CZMG agents becomes a
%monotonically increasing function of $f1$. 

%We observe that CZMG agents having memory length $m1=1$ do not have
%any advantage over the DIMG agents, but as $m1$ is increased they show
%a relatively better performance for an optimal range of population
%fraction $f1$ (for simplicity, we keep the memory size of the DIMG
%agents, $m_2$, fixed to 1).
Let us now consider the performance of the Type 2 agents. When playing
against Type 1 agents with low memory length $m_1$, Type 2 agents
achieve their highest payoff when $f_1 \rightarrow 1$, i.e., when they
are present in extremely small numbers in the population. 
In other words, to achieve the best performance out of availability of
detailed data, it is important to have the size of the group to which this
data is available as small as possible in this regime of low $m_1$.
As more agents have access to this data (i.e., decreasing $f_1$),
their payoff is eroded until they actually perform worse
than those having coarser-grained data, i.e., Type 1 agents.
Thus, access
to more and better data is not by itself a determining factor for success in
a complex adaptive situation.

%Switchover of best performance of Type 2 agents to low $f_1$. Now
%Type 1 agents have quantitatively more information, but Type 2 agents
%can offset it when present in large numbers by coordination effects
As the memory length $m_1$ of the Type 1 agents increases, the optimal
population fraction at which Type 2 agents achieve the highest payoff
decreases from the neighborhood of $f_1=1$. In fact, in the case of Type 2
agents having memory length $m_2 = 2$ (the optimal memory length for a
population exclusively composed of such agents), their best
performance is achieved as
$f_1 \rightarrow 0$. Thus, in this high $m_1$ regime ($m_1 \geq
6$ for the case of $m_2=2$), Type 2 agents achieve high payoffs by 
dominating the population. By contrast, Type 1 agents
do better than Type 2 agents for large $f_1$ as a result of emergent
coordination within their group. Indeed, in this regime, for any given
$m_1$ the payoff of Type 1 agents increases with $f_1$. 
%In other words, one or a few DIMG
%agents will not perform very well when facing CZMG agents with
%sufficiently large memory size $m1$.
Thus, the outcome is not symmetric for agents having access to
information at the two extreme levels of coarse-graining.
%If you are DIMG having lower information in bits compared to CZMG
%agents (i.e., high m1), it's not advantageous to be very few (i.e.,
%better performance of DIMG agentys occur at low f1 where they are
%larger in number), while if you are a CZMG agent having lower
%information in bits compared to DIMG agents (i.e., low m1) it is
%advantageous to be present in small numbers (i.e., CZMG agents have
%better payoffs for f1 going to 0
Note that as $m_1$ is increased more and more, the strategy space for
Type 1 agents become so large that the action of the agents
essentially resemble randomly choosing between the two options.
If $m_2$ is also sufficiently large ($>2$), both types of agents
achieve similar payoffs, equal to that obtained by a random choice
strategy (see Supplementary Information).
%As $m_1$ becomes larger, the agents become closer to random
%especially at low f_1 end

%\section{Discussion and Conclusions}
%\label{sec4}
To conclude, we have shown that information asymmetry among agents in
a complex adaptive system can have surprising consequences.
Specifically, in a
system where agents compete for a limited resource using strategies
based on information about the collective behavior in previous
interactions, asymmetry arising from individuals having
access only to data coarse-grained to different levels can result in
agents with more and better data performing worse than others under
certain circumstances. 
Such counter-intuitive effects arise from
predictable patterns emerging in the collective information about the
system at a certain level of coarse-graining and thus discernible only
to agents privy to that level. This provides them a
competitive advantage when the population is dominated
by agents of a different type who do not have access to the
coarse-graining level at which such
patterns generated by their own collective activity are apparent.
The relation between the relative performance of the different types
of agents and the nature of information asymmetry is therefore
crucially dependent on the exact composition of the population to
which they belong. 
%This paradigm of information usage at different levels of
%coarse-graining by different agents is quite general and the
%collective behavior thereof is quite general and would be worth
%looking in the broader context of strategic interactions of different kinds
Our results imply that striving to acquire and process ever increasing
quantities of data in the
hope of making more accurate predictions in complex adaptive systems,
such as financial markets, may sometimes be counter-productive.
%%NEW
While concerns about the potential pitfalls of ``big data'' have been
voiced earlier~\cite{silver2012}, we provide possibly the first rigorous
demonstration using a quantitative model of how such a failure can come about.
%%NEW
The insights gained from our study are quite general and should apply
%extends beyond the setting of competition for a limited resource to 
to the broader context of
strategic interactions between a large number of adaptive agents.

%MOVE TO DISCUSSION
%The MG paradigm was itself inspired by the El Farol Bar problem 
%where a given number of agents compete for a
%limited resource~\cite{arthur1994}. 
%
%Here we have considered two extreme cases of information
%coarse-graining. In certain situations, it may seem more natural to
%look at very coarse-grained data, e.g., when deciding on which of the
%two routes to choose during one's commute. One may be satisfied with
%data which only states which route took the longer time to traverse
%in the past few days to take a decision rather than trying to
%discover precisely how many commuters chose each route on each of the
%days. On the other hand, in other situations, one may prefer detailed
%data. For example, when investing in a stock, one may not be
%satisfied to know only whether the price rose or fell in the past few
%days (i.e., only the direction of the price fluctuations), but rather
%by how much amount it changed on each day (the magnitude). It may
%appear that one can take a more nuanced decision based on the more
%fine-grained data. In general, there will be many other levels of
%coarse-graining possible in between these two extremes and one can
%ask whether there is an optimal level of coarse-graining of
%information that will confer an advantage in a specific circumstance.
%This issue of the possible role of information granularity on the
%relative performance of agents will be addressed in a future work.

\acknowledgments
We would like to thank Shakti N. Menon for helpful discussions.
This work was supported in part by the IMSc Econophysics project (XII
Plan) funded by the Department of Atomic Energy, Government of India.
We thank IMSc for providing access to the supercomputing cluster
``Satpura'', which is partially funded by DST.

\pagebreak
\begin{table*}
{\large \bf SUPPLEMENTARY INFORMATION}
\end{table*}
\setcounter{figure}{0}
\renewcommand\thefigure{S\arabic{figure}}
\begin{figure*}
\begin{center}
\includegraphics[width=.45\linewidth]{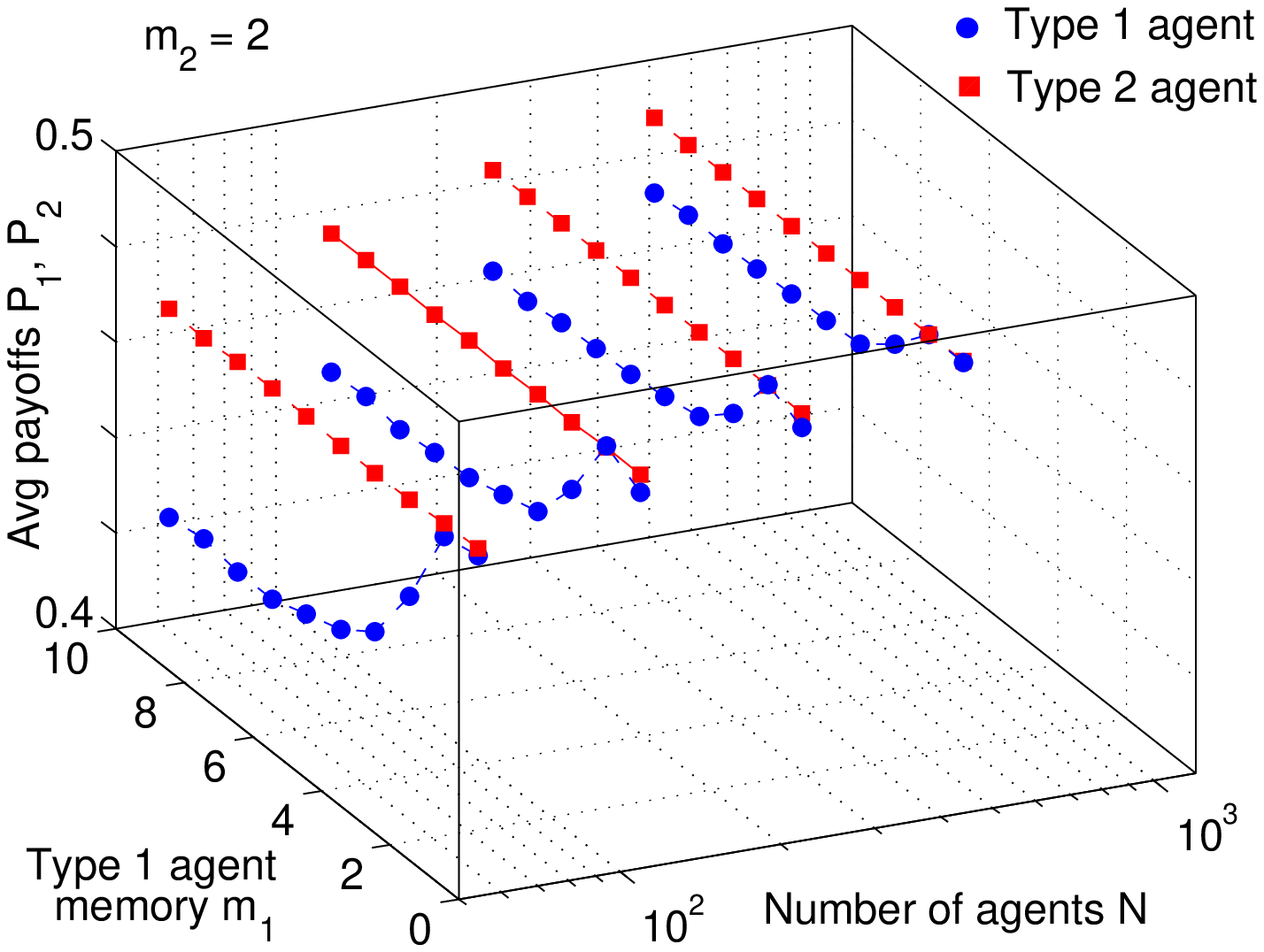}
\end{center}
\caption{Average payoffs $P_1$ , $P_2$ of Type 1 and Type 2 agents
(respectively) shown as a function of the memory length $m_1$
of a single Type 1 agent interacting with $N-1$ Type 2 agents
with memory length $m_2 = 2$ for different population sizes N.
Note that when the lone Type 1 agent does not have any relative
advantage over the Type 2 agents. However, it still receives its
highest payoff when $m_1 = 2$ (as in Fig.~\ref{fig:3} for $m_2 = 1$).
Payoffs are averaged over $10^4$ iterations in the steady state and
over $100$ different realizations.
}
\label{fig:s0}
\end{figure*}

\begin{figure*}
\begin{center}
\includegraphics[width=.45\linewidth]{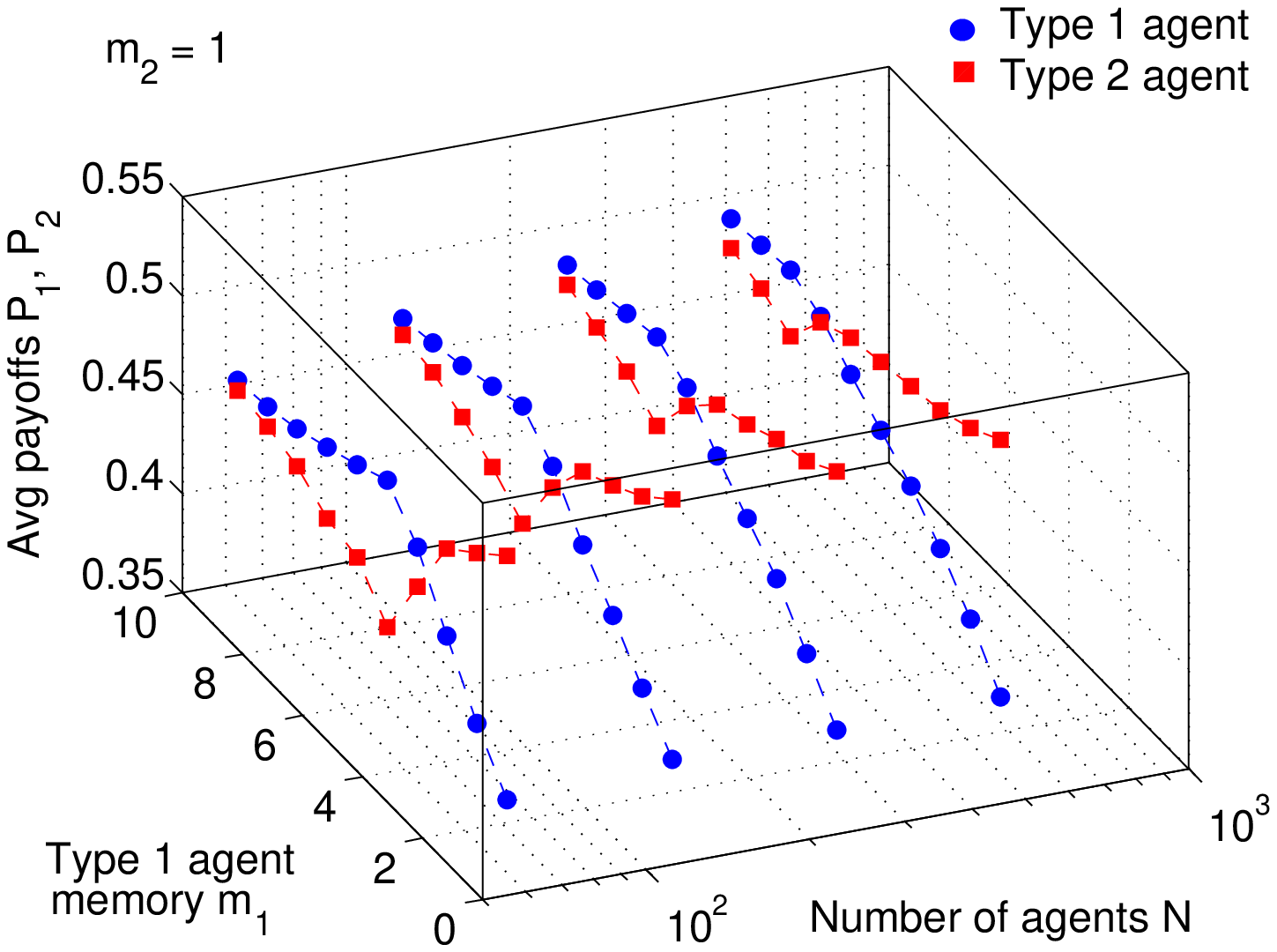}
\end{center}
\caption{Average payoffs $P_1, P_2$ of Type 1 and Type 2 agents
(respectively) shown as a function of the memory length $m_1$ of a
population of $N-1$ Type 1 agents interacting with a single Type 2
agent with memory length $m_2 = 1$ for different values of $N$.
Note that the lone Type 2 agent enjoys a significant advantage over
the rest of the population for low values of $m_1$. The trend of the
payoff of the Type 2 agent as a function of $m_1$ appears to mirror
that of the Type 1 agents.
Payoffs are averaged over $10^4$ iterations in the steady state and
over 100 different realizations.
A similar profile is seen when the memory length of the Type 2 agents
is $m_2 = 2$.
}
\label{fig:s1}
\end{figure*}

\begin{figure*}
\begin{center}
\includegraphics[width=.45\linewidth]{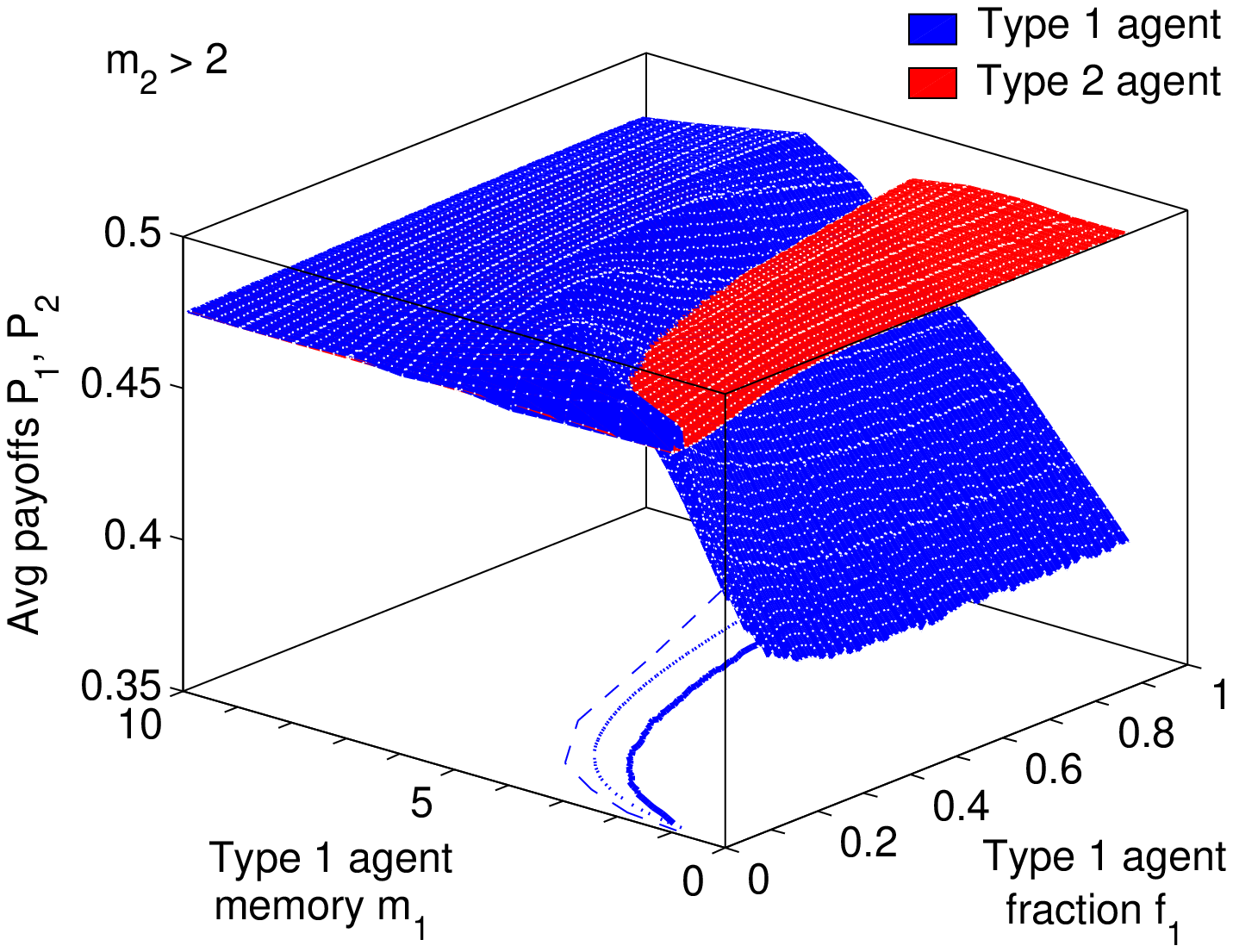}
\end{center}
\caption{The average payoffs $P_1, P_2$ of Type 1 (shown in blue) and
Type 2 agents (red) comprising a population of size $N$ (=255) for
different population fractions $f_1$ and memory length $m_1$ of Type 1
agents. As Type 2 agents with sufficiently large memory length ($m_2 >
2$) effectively use random choice strategy~\cite{sasidevan}, here the
Type 2 agents are assumed to be randomly choosing between the two
possible options. The contours separate the regions
in the ($m_1, f_1$) parameter space where Type 1 agents have a
relative advantage over Type 2 agents and vice versa. 
The broken curve represents the optimal population fraction $f_1^*$
of Type 1 agents with a given memory length $m_1$ at which they
receive the highest payoff. The dotted curve is the value of 
$m_1$ at which $N f_1$ Type 1 agents are expected to have maximum
payoff in absence of any Type 2 agents. Payoffs are averaged
over $10^4$ iterations in the steady state and over $100$ different
realizations.
}
\label{fig:s2}
\end{figure*}

\end{document}